\documentclass[nofootinbib,floatfix,amsmath,amssymb,showpacs,superscriptaddress,notitlepage]{revtex4-1}

\usepackage{rotating}
\usepackage{graphicx}
\usepackage{dcolumn}
\usepackage{bm}
\usepackage{color}
\usepackage{multirow}

\newcommand{\omc}{{\Omega_c}}
\newcommand{\omcs}{{\Omega_{c}^\ast}}
\newcommand{\oog}{\Omega_c \gamma \rightarrow\Omega_c^\ast}
\newcommand{\ogo}{\Omega_c^\ast \rightarrow\Omega_c \gamma }
\newcommand{\ojo}{\Omega_c^\ast j^\mu \Omega_c}

\begin{document}

\title{$\Omega_c \gamma \rightarrow\Omega_c^\ast$ transition in lattice QCD}

\author{H. Bahtiyar}
\affiliation{Department of Physics, Yildiz Technical University, Davutpasa Campus Esenler 34210 Istanbul
Turkey}
\affiliation{Department of Physics, Mimar Sinan Fine Arts University, Bomonti 34380 Istanbul Turkey}
\author{K. U. Can}
\affiliation{Department of Physics, H-27, Tokyo Institute of Technology, Meguro, Tokyo 152-8551 Japan}
\author{G. Erkol}
\affiliation{Department of Natural and Mathematical Sciences, Faculty of Engineering, Ozyegin University, Nisantepe Mah. Orman Sok. No:34-36, Alemdag 34794 Cekmekoy, Istanbul Turkey}
\author{M. Oka}%
\affiliation{Department of Physics, H-27, Tokyo Institute of Technology, Meguro, Tokyo 152-8551 Japan}
\affiliation{Advanced Science Research Center, Japan Atomic Energy Agency, Tokai, Ibaraki, 319-1195 Japan}

\date{\today}

\begin{abstract}
 We study the electromagnetic $\Omega_c \gamma \rightarrow\Omega_c^\ast$ transition in 2+1 flavor lattice QCD, which gives access to the dominant decay mode of $\Omega_c^\ast$ baryon. The magnetic dipole and the electric quadrupole transition form factors are computed. The magnetic dipole form factor is found to be mainly determined by the strange quark and the electric quadrupole form factor to be negligibly small, in consistency with the quark model. We also evaluate the helicity amplitudes and the decay rate.

\end{abstract}
\pacs{14.20.Lq, 12.38.Gc, 13.40.Gp }
\keywords{charmed baryons, electric and magnetic form factor, lattice QCD}
\maketitle

\section{Introduction}
Recently there has been a significant progress in our understanding of the heavy-flavor hadron sector. Experimentally, all the ground-state single-charmed baryons and several excited states, as predicted by the quark model, have been confirmed~\cite{Agashe:2014kda}. Unlike other single-charmed baryons, a precise observation of $\omc$ baryon had been long overdue. Only recently, Belle Collaboration has made a rigorous experimental study of $\omc$ using the decay $\omc^0\rightarrow \Omega^-\pi^+$~\cite{Solovieva:2008fw}.

The $\Omega_c^0 (css)$ has the quantum numbers $J^P=\frac{1}{2}^+$ and is the heaviest known single charmed hadron that decays weakly. Within the multiplet structure of flavor SU(4), $\omc$ belongs to a sextet of flavor symmetric states, which sits on the second layer of the flavor mixed-symmetric 20-plet. The average mass value reported by the Particle Data Group (PDG) is $m_{\Omega_c}=2695.2\pm 1.7$~MeV~\cite{Agashe:2014kda}.

The excited $\omcs^0(css)$ baryon was first observed by BABAR Collaboration in the radiative decay $\ogo$~\cite{Aubert:2006je}. Belle Collaboration has confirmed their observation by reconstructing $\omcs$ in the same radiative decay mode~\cite{Solovieva:2008fw}. They measured the relative mass difference with respect to the ground state $m_\omcs-m_\omc=70.7\pm 0.9^{+0.1}_{-0.9}$~MeV in very good agreement with the BABAR observation. The average mass value reported by PDG is $m_{\omcs}=2765.9 \pm 2.0$~MeV~\cite{Agashe:2014kda}. The quantum numbers have not been measured but natural assignment is that it completes the ground state $J^P=\frac{3}{2}^+$ sextet, which sits on the second layer of the flavor-symmetric 20-plet of SU(4). The mass difference with respect to the ground state is too small for any strong decay to occur, therefore the radiative channel $\ogo$ is the dominant decay mode.  

In addition to BABAR and Belle Collaborations, experimental facilities such as LHCb, PANDA, Belle II, BESIII and J-PARC are expected to give a more precise determination of charmed baryon spectroscopy. Concurrently, recent lattice-QCD studies on the spectroscopy of charmed hadrons are also very promising. The ground-state charmed baryons with spin up to 3/2 have been studied in quenched~\cite{Lewis:2001iz,Mathur:2002ce} and full QCD~\cite{Namekawa:2013vu, Alexandrou:2012xk, Briceno:2012wt,Alexandrou:2014sha}. The results for baryon masses as determined from lattice QCD are in good agreement with experiment.

Recently we have examined the charmed baryons in lattice QCD in order to reveal their electromagnetic structure~\cite{Can:2013zpa, Can:2013tna}. We have extracted the charge radii and magnetic moments of $J=1/2$ charmed baryons by computing their elastic electromagnetic form factors on the lattice. A similar study for $J=3/2$ baryons is in progress. A phenomenologically more interesting problem is the electromagnetic transitions between $J=1/2$ and $J=3/2$ baryons, which are more accessible by experiments as explained above.

Being motivated by the experimental discovery of the $\omcs$ baryon in the radiative decay mode, in the present work we give a timely study of the $J=1/2 \rightarrow J=3/2$ electromagnetic transition of single charmed strange baryons in lattice QCD. In particular we study the $\oog$ transition, which gives access to three Sachs form factors, the helicity amplitudes, the decay width and the lifetime. This work is reminiscent of Refs.~\cite{Leinweber:1992pv, Alexandrou:2003ea, Alexandrou:2004xn, Alexandrou:2007dt}, where the electromagnetic $N$ to $\Delta$, and the other octet to decuplet transitions have been studied. The electromagnetic transitions of charmed baryons have also been studied within heavy hadron chiral perturbation theory~\cite{Cheng:1992xi, Banuls:1999br,Jiang:2015xqa} and in the quark models~\cite{Dey:1994qi, Ivanov:1996fj, Ivanov:1999bk}. 

The three transition form factors, namely, the magnetic dipole ($M1$), the electric quadrupole ($E2$) and the electric charge quadrupole ($C2$) provide valuable information about the structure and shape of $J=1/2$ and $J=3/2$ baryons. Earlier studies have focused on the transition moments between $N$ and $\Delta$. Experimentally, pure single spin-flip $M1$ transition has been found to dominate. Of special interest is the small but non-vanishing values of $E2$ and $C2$ moments, implying that the shapes of $N$ and $\Delta$ deviate from spherical symmetry~\cite{Buchmann:2001gj}. Quark models predict a nonzero value for $E2$ and $C2$~\cite{Isgur:1981yz}, which has also been confirmed experimentally~\cite{Mertz:1999hp, Joo:2001tw}. However, the results from various theoretical approaches are not in complete agreement with experiment and this issue is still unsettled.

The experimental results for $\oog$, on the other hand, are not yet precise enough to allow a determination of the transition strengths. In this work, we will mainly focus on the $M1$ and $E2$ transition form factors. Unlike in the case of $N\gamma\rightarrow\Delta$, the mass splitting between $\omc$ and $\omcs$ can be reproduced on the lattice and an accurate contact can be made to phenomenological observables via these two form factors. We employ near physical 2+1-flavor lattices that correspond to a pion mass of approximately 156~MeV. The data for electromagnetic transition form factors are often noisier than those for elastic form factors, particularly for $C2$ form factor. Considering also the limited number of gauge configurations we have at the smallest quark mass, we study the $M1$ and $E2$ form factors for the lowest allowed lattice momentum transfer. We, however, make contact with the transition moments at zero-momentum transfer by assuming a simple scaling at low momentum transfer values~\cite{Leinweber:1992pv}.  

\section{Lattice formulation}
Electromagnetic transition form factors for $\oog$ can be calculated by considering the baryon matrix elements of the electromagnetic vector current $\mathcal{J}_\mu=  \sum \limits_{q}{} \frac{2}{3} \bar{c}(x) \gamma_{\mu} c(x)-\frac{1}{3} \bar{s}(x) \gamma_{\mu} s(x)$. The matrix element can be written in the following form
	\begin{equation}\label{matel}
	\langle \omcs (p^\prime,s^\prime)|\mathcal{J}_\mu|\omc(p,s)\rangle= i \sqrt{\frac{2}{3}}\left(\frac{m_\ast \
m}{E_\ast({\bf p^\prime})E({\bf p})}\right)\bar{u}_\tau(p^\prime,s^\prime) {\cal O}^{\tau\mu} u(p,s),
	\end{equation}
with the operator ${\cal O}^{\tau\mu}$ given in terms of Sachs form factors as~\cite{Jones:1972ky}
	\begin{equation}\label{sachs}
	{\cal O}^{\tau\mu}=G_{M1}(q^2) K_{M1}^{\tau\mu}+G_{E2}(q^2) K_{E2}^{\tau\mu}+G_{C2}(q^2) K_{C_2}^{\tau\mu},
	\end{equation}
where
	\begin{align}
		K_{M1}^{\tau\mu}&=-3\Big((m_\ast+m)^2-q^2\Big)^{-1}i\epsilon^{\tau\mu}(Pq)~(m_\ast+m)/2m,\\
		K_{E2}^{\tau\mu}&=-K_{M1}^{\tau\mu}-6\Omega^{-1}(q^2)~ i\epsilon^{\tau\beta}(Pq)~ \epsilon^{\mu\beta}(p^\prime q)~\gamma_5 (m_\ast+m)/m,\\
		K_{C_2}^{\tau\mu}&=-3\Omega^{-1}(q^2)~q^\tau (q^2 P^\mu-q\cdot P~ q^\mu)~i\gamma_5(m_\ast+m)/m.
	\end{align}
Here $p$ and $p^\prime$ denote the incoming and the outgoing momenta, respectively, $q=p^\prime-p$ is the transferred four\--momentum, $P=(p^\prime+p)/2$ and 
	\begin{equation}
		\Omega(q^2)=\Big((m_\ast+m)^2-q^2\Big)\Big((m_\ast-m)^2-q^2\Big).
	\end{equation}
We use the shorthand notation $\epsilon^{\tau\mu}(Pq)=\epsilon^{\tau\mu\alpha\nu}P^\alpha q^\nu$. The spins are denoted by $s$, $s^\prime$ and the masses of $\omcs$ and $\omc$ by $m_\ast$ and $m$, respectively. $u(p,s)$ is the Dirac spinor and $u_\tau(p,s)$ is the Rarita-Schwinger spin vector. For real photons, $G_{C2}(0)$ does not play any role as it is proportional to the longitudinal helicity amplitude.

The Rarita-Schwinger spin sum for the spin-3/2 field in Euclidean space is given by
\begin{align}
	\sum_s u_\sigma(p,s)& \bar{u}_\tau(p,s)=\frac{-i\gamma\cdot p+m_\ast}{2m_\ast}\left[g_{\sigma\tau}-\frac{1}{3}\gamma_\sigma \gamma_\tau+\frac{2p_\sigma p_\tau}{3m_\ast^2}-i\frac{p_\sigma \gamma_\tau-p_\tau \gamma_\sigma}{3m_\ast}\right],
\end{align}
and the Dirac spinor spin sum by
\begin{equation}
	\sum_s u(p,s) \bar{u}(p,s)=\frac{-i\gamma\cdot p +m}{2m}.
\end{equation}
We refer the form factors $G_{M1}$, $G_{E2}$ and $G_{C2}$ as the magnetic dipole, the electric quadrupole and the electric charge quadrupole transition form factors, respectively. 

\begin{widetext}
To extract the form factors we consider the following matrix elements, 
\allowdisplaybreaks{
\begin{align}
	\begin{split}\label{deltacf}
	&\langle G_{\sigma\tau}^{\omcs\omcs}(t; {\bf p};\Gamma_4)\rangle=\sum_{\bf x}e^{-i{\bf p}\cdot {\bf x}}\Gamma_4^{\alpha\alpha^\prime} \times \langle \text{vac} | T [\eta_{\sigma}^\alpha(x) \bar{\eta}_{\tau}^{\alpha^\prime}(0)] | \text{vac}\rangle,
	\end{split}\\
	\begin{split}\label{nuccf}
	&\langle G^{\omc\omc}(t; {\bf p};\Gamma_4)\rangle=\sum_{\bf x}e^{-i{\bf p}\cdot {\bf x}}\Gamma_4^{\alpha\alpha^\prime} \times \langle \text{vac} | T [\eta^\alpha(x) \bar{\eta}^{\alpha^\prime}(0)] | \text{vac}\rangle,
	\end{split}\\
	\begin{split}\label{thrpcf}
	&\langle G_\sigma^{\ojo}(t_2,t_1; {\bf p}^\prime, {\bf p};\mathbf{\Gamma})\rangle=-i\sum_{{\bf x_2},{\bf x_1}} e^{-i{\bf p}\cdot {\bf x_2}} e^{i{\bf q}\cdot {\bf x_1}} \mathbf{\Gamma}^{\alpha\alpha^\prime} \langle \text{vac} | T [\eta_\sigma^\alpha(x_2) j_\mu(x_1) \bar{\eta}^{\alpha^\prime}(0)] | \text{vac}\rangle,
	\end{split}
\end{align}
}%
with the spin projection matrices defined as 
\begin{equation}
	\Gamma_i=\frac{1}{2}\left(\begin{matrix}\sigma_i & 0 \\ 0 & 0 \end{matrix}\right), \qquad \Gamma_4=\frac{1}{2}\left(\begin{matrix}I & 0 \\ 0 & 0 \end{matrix}\right).
\end{equation}
Here, $\alpha$, $\alpha^\prime$ are the Dirac indices, $\sigma$ and $\tau$ are the Lorentz indices of the spin-3/2 interpolating field and $\sigma_i$ are the Pauli spin matrices. An initial $\omc$ state is created at time zero and interacts with the external electromagnetic field at time $t_1$. At time $t_2$ the final $\omcs$ state is annihilated.

The baryon interpolating fields are chosen, similarly to those of $\Delta$ and $N$ as
\begin{align}
		&\eta_\mu(x)=\frac{1}{\sqrt{3}}\epsilon^{ijk} \left\{2[s^{T i}(x) C \gamma_\mu c^j(x)]s^k(x)+[s^{T i}(x) C \gamma_\mu s^j(x)]c^k(x)\right\},\label{deltaint}\\
		&\eta(x)=\epsilon^{ijk}[s^{T i}(x) C \gamma_5 c^j(x)]s^k(x),
\end{align}
where $i$, $j$, $k$ denote the color indices and $C=\gamma_4\gamma_2$. It has been shown in Ref.~\cite{Alexandrou:2014sha} that the interpolating field in Eq.~\eqref{deltaint} has minimal overlap with spin-1/2 states and therefore does not need any spin-3/2 projection.

To extract the form factors, we calculate the following ratio of the two- and three-point functions:
\begin{equation}\label{ratio}
	R_\sigma(t_2,t_1;{\bf p}^\prime,{\bf p};\mathbf{\Gamma};\mu)=
	\cfrac{\langle G_\sigma^{\ojo}(t_2,t_1; {\bf p}^\prime, {\bf p};\mathbf{\Gamma})\rangle}{\langle \delta_{ij} G_{ij}^{\omcs\omcs}(t_2; {\bf p}^\prime;\Gamma_4)\rangle} \left[\cfrac{ \delta_{ij} G_{ij}^{\omcs\omcs} (2t_1; {\bf p}^\prime;\Gamma_4)\rangle }{ G^{\omc\omc} (2t_1; {\bf p};\Gamma_4)\rangle }\right]^{1/2}.
\end{equation} 
In the large Euclidean time limit, $t_2-t_1$ and $t_1\gg a$, the unknown normalization factors cancel and the time dependence of the correlators can be eliminated. Then the ratio in Eq.~(\ref{ratio}) reduces to the desired form
\begin{equation}\label{desratio}
	R_\sigma(t_2,t_1;{\bf p^\prime},{\bf p};\Gamma;\mu)\xrightarrow[t_2-t_1\gg a]{t_1\gg a} \Pi_\sigma({\bf p^\prime},{\bf p};\Gamma;\mu),
\end{equation}
where we can look for a plateau to extract the form factors. We choose the ratio in Eq.~\eqref{ratio} among several other alternatives used in the literature~\cite{Leinweber:1992pv, Alexandrou:2003ea, Alexandrou:2004xn, Alexandrou:2007dt} due to the good plateau region and the quality of the signal it yields.

We single out the Sachs form factors by choosing appropriate combinations of Lorentz direction $\mu$ and projection matrices $\Gamma$. When $\omc$ is produced at rest and momentum is inserted in one spatial direction, we have~\cite{Alexandrou:2003ea}
\begin{align}
	\begin{split}\label{cff}
	G_{C2}(q^2)&=C(\mathbf{q}^2)\frac{2m_\ast}{\mathbf{q}^2} \Pi_k({\bf q},{\bf 0};i \Gamma_k;4)
	\end{split}
	\\
	\begin{split}\label{mff}
	G_{M1}(q^2)&=C(\mathbf{q}^2)\frac{1}{|\mathbf{q}|}\left[\Pi_l(q_k,{\bf 0}; \Gamma_k;l)-\frac{m_\ast}{E_\ast}\Pi_k(q_k,{\bf 0}; \Gamma_l;l)\right], 
	\end{split}\\
	\begin{split}\label{qff}
	G_{E2}(q^2)&=C(\mathbf{q}^2)\frac{1}{|\mathbf{q}|}\left[\Pi_l(q_k,{\bf 0}; \Gamma_k;l)+\frac{m_\ast}{E_\ast}\Pi_k(q_k,{\bf 0}; \Gamma_l;l)\right],  
	\end{split}
\end{align}
where
\begin{equation}
	C(\mathbf{q^2})=2\sqrt{6}\frac{E_\ast m_\ast}{m+m_\ast}\left( 1+\frac{m_\ast}{E_\ast}\right)^{1/2} \left(1+\frac{\mathbf{q}^2}{3m_\ast^2} \right)^{1/2},
\end{equation}
and $k$ and $l$ are two distinct indices running from 1 to 3. When $\omcs$ is produced at rest, $m_\ast=E_\ast$ in Eqs.~(\ref{cff}-\ref{qff}) and 
\begin{equation}
	C(\mathbf{q^2})= 2\sqrt{6} \frac{E m}{m_\ast+m}\left( 1+\frac{m}{E}\right)^{1/2} \left(1+\frac{\mathbf{q}^2}{3m_\ast^2} \right)^{1/2}.
\end{equation}

\end{widetext}

We have run our simulations on gauge configurations generated by PACS-CS collaboration~\cite{Aoki:2008sm} with the nonperturbatively $O(a)$-improved Wilson quark action and the Iwasaki gauge action. The details of the gauge configurations are given in Table~\ref{lat_det}. The simulations are carried out with near physical $u$,$d$ sea quarks of hopping parameter $\kappa^{u,d}=$ 0.13781. This corresponds to a pion mass of approximately 156~MeV~\cite{Aoki:2008sm}. The hopping parameter for the sea $s$ quark is fixed to $\kappa_\text{sea}^{s}=0.13640$ and the hopping parameter for the valence $s$-quark is taken to be the same.

%%%%%%%%% lattice details %%%%%%%%%%%%%%%%%%%%%%
\begin{table}[ht]
	\caption{ The details of the gauge configurations used in this work~\cite{Aoki:2008sm}. We list the number of flavors~($N_f$), the lattice spacing~($a$), the lattice size~($L$), inverse gauge coupling~($\beta$), clover coefficient~($c_{SW}$), number of gauge configurations employed and the corresponding pion mass~($m_\pi$).
}
%	\addtolength{\tabcolsep}{6pt}
\begin{center}
	{
	\setlength{\extrarowheight}{7pt}
\begin{tabular*}{0.8\textwidth}{@{\extracolsep{\fill}}cccccccc}
			\hline\hline 
			$N_s^3\times N_t$  & $N_f$ & $a$~[fm] &  $L$~[fm] & $\beta$ & $c_{SW}$ & $\#$ of conf.  & $m_\pi$~[MeV]\\
			\hline
			$32^3 \times 64$ & 2+1 & 0.0907(13)  & 2.90 & 1.90 & 1.715 & 194 & 156(7)(2)\\
			\hline \hline			
\end{tabular*}
	\label{lat_det}
	}
\end{center}
\end{table}
%%%%%%%%%%%%%%%%%%%%%%%%%%%%%%%%%%%%%%%%%%%%%%%%%%%%%%%%%%%%%%%%%%

As we perform the simulations at only one (near-physical) quark mass, a chiral extrapolation cannot be made. However, we can make an estimation of an uncertainty anticipated from such an extrapolation, based on our simulations of elastic $\Omega_c$ electromagnetic form factors. We have performed the chiral extrapolations for electric/magnetic charge radii and the magnetic moment of $\Omega_c$ baryon in Ref.~\cite{Can:2013tna} again, including the data at $m_\pi\simeq 156$~MeV. We tried constant, linear and quadratic fit functions. For all cases, the chiral-extrapolated values and those at the smallest pion mass are in very good agreement within their error bars. Different fit forms we use imply a systematic error of less than 1\%. Hence, we anticipate to have a similarly negligible error from such an extrapolation of $M1$ and $E2$ form factors here.

For the charm quarks, we apply the Fermilab method~\cite{ElKhadra:1996mp} in the form employed by the Fermilab Lattice and MILC Collaborations~\cite{Burch:2009az, Bernard:2010fr}. A similar approach has been recently used to study charmonium, heavy-light meson resonances and their scattering with pion and kaon~\cite{Mohler:2011ke, Mohler:2012na, Mohler:2013rwa}. In this simplest form of the Fermilab method, the Clover coefficients $c_E$ and $c_B$ in the action are set to the tadpole-improved value $1/u_0^3$, where $u_0$ is the average link. Following the approach in Ref.~\cite{Mohler:2011ke}, we estimate $u_0$ to be the fourth root of the average plaquette. We determine the charm-quark hopping parameter $\kappa_c$ nonperturbatively. To this end, we measure the spin-averaged static masses of charmonium and heavy-light mesons and tune their values accordingly to the experimental results, which yields $\kappa_c=0.1246$~\cite{Can:2013tna}.

We make our simulations with the lowest allowed lattice momentum transfer $q=2\pi/N_s a$, where $N_s$ is the spatial dimension of the lattice and $a$ is the lattice spacing. This corresponds to three-momentum squared value of ${\bf q}^2=0.183$~GeV$^2$. In order to access the values of the form factors at $Q^2=-q^2=0$, we will apply the procedure in Ref.~\cite{Leinweber:1992pv} and assume that the momentum-transfer dependence of the transition form factors is the same as the momentum dependence of the $\omcs$ baryon charge form factor. Such a scaling is also suggested by the experimentally measured proton form factors and it was used in previous analyses such as baryon octet to decuplet electromagnetic transition form factors~\cite{Leinweber:1992pv}. While extrapolations in finite momentum suffer from large statistical errors since one has to rely on a functional form, the scaling approach provides a more precise determination of the form-factor values at zero momentum transfer. In applying this procedure, we consider $s$ and $c$ quark sectors separately as their contributions to the charge form factors scale differently. For instance, the scaling of $G_{M1}$ is given by
\begin{equation}\label{scaling}
	G_{M1}^{s,c}(0)=G_{M1}^{s,c}(q^2)\frac{G_{E}^{s,c}(0)}{G_{E}^{s,c}(q^2)}.
\end{equation} 
The heavy-quark contribution yields a harder form factor whereas the light-quark contribution is soft and the form factor falls off more rapidly~\cite{Can:2013zpa, Can:2013tna}. The form factors are extracted in two kinematically different cases. In the first case, the $\omcs$ is produced at rest and the $\omc$ has momentum $-{\bf q}$ and in the second case, the $\omc$ is at rest and $\omcs$ carries momentum ${\bf q}$. 

In order to increase statistics, we insert positive and negative momentum in one of the spatial directions and make a simultaneous fit over all available data. We also consider current along all spatial directions. The source-sink time separation is fixed to 1.09 fm ($t_2=12a$), which has been shown to be sufficient to avoid excited state contaminations for electromagnetic form factors~\cite{Can:2013tna}. Using translational symmetry, we have employed multiple source-sink pairs by shifting them 12 lattice units in the temporal direction. All statistical errors are estimated by the single-elimination jackknife analysis. We consider point-split lattice vector current
\begin{equation}
j_\mu = 1/2[\bar{q}(x+\mu)U^\dagger_\mu(1+\gamma_\mu)q(x) -\bar{q}(x)U_\mu(1-\gamma_\mu)q(x+\mu)],
\end{equation}
which is conserved by Wilson fermions.

A wall-source/sink method~\cite{Can:2012tx} has been employed, which provides a simultaneous extraction of all spin, momentum and projection components of the correlators. The gauge non-invariant wall source/sink requires fixing the gauge. We fix the gauge to Coulomb, which gives a somewhat better coupling to the ground state. The delta function operator is smeared over the three spatial dimensions of the time slice where the source is located, in a gauge-invariant manner using a Gaussian form. In the case of $s$ quark, we choose the smearing parameters so as to give a root-mean-square radius of $\langle r_l \rangle \sim 0.5$~fm. As for the charm quark, we adjust the smearing parameters to obtain $\langle r_c \rangle=\langle r_l \rangle/3$.

%%%%%%%%% baryon masses %%%%%%%%%%%%%%%%%%%%%%
\begin{table}[ht]
	\caption{ The $\omc$ and $\omcs$ masses (at a pion mass of $m_\pi=156$~MeV) together with the experimental values~\cite{Agashe:2014kda} and those obtained by PACS-CS~\cite{Namekawa:2013vu} (at the physical point). We have also included results by ETMC~\cite{Alexandrou:2014sha} and Briceno \emph{et al.}~\cite{Briceno:2012wt} obtained by chiral extrapolation. 
}
%	\addtolength{\tabcolsep}{6pt}
\begin{center}
	{
	\setlength{\extrarowheight}{7pt}
\begin{tabular*}{0.8\textwidth}{@{\extracolsep{\fill}}c|ccccc}
			\hline\hline 
			  & This work & PACS-CS~\cite{Namekawa:2013vu} &  ETMC~\cite{Alexandrou:2014sha} & Briceno \emph{et al.}~\cite{Briceno:2012wt} & Exp.~\cite{Agashe:2014kda}   \\
			\hline \hline
			$m$~[GeV] & 2.750(15) & 2.673(17)  & 2.629(22) & 2.681(48) & 2.695(2)\\
			$m_\ast$~[GeV] & 2.828(15) & 2.738(17)  & 2.709(26) & 2.764(49) & 2.766(2)\\
			\hline \hline			
\end{tabular*}
	\label{bar_mass}
	}
\end{center}
\end{table}
%%%%%%%%%%%%%%%%%%%%%%%%%%%%%%%%%%%%%%%%%%%%%%%%%%%%%%%%%%%%%%%%%%

\section{Numerical results and discussion}
We extract the $\omc$ and $\omcs$ masses using the two-point correlators in Eqs.~\eqref{deltacf} and \eqref{nuccf}. Our results for the $\omc$ and $\omcs$ masses are given in Table~\ref{bar_mass}, together with the experimental values and those obtained by other lattice collaborations. While we see a few percent discrepancy between our results obtained at a pion mass of $m_\pi=156$~MeV and those of PACS-CS obtained at the physical point, the mass splitting $m_\ast-m$ is nicely produced in agreement with experiment. 

\begin{figure}[t]
	\centering
	\includegraphics[width=0.9\textwidth]{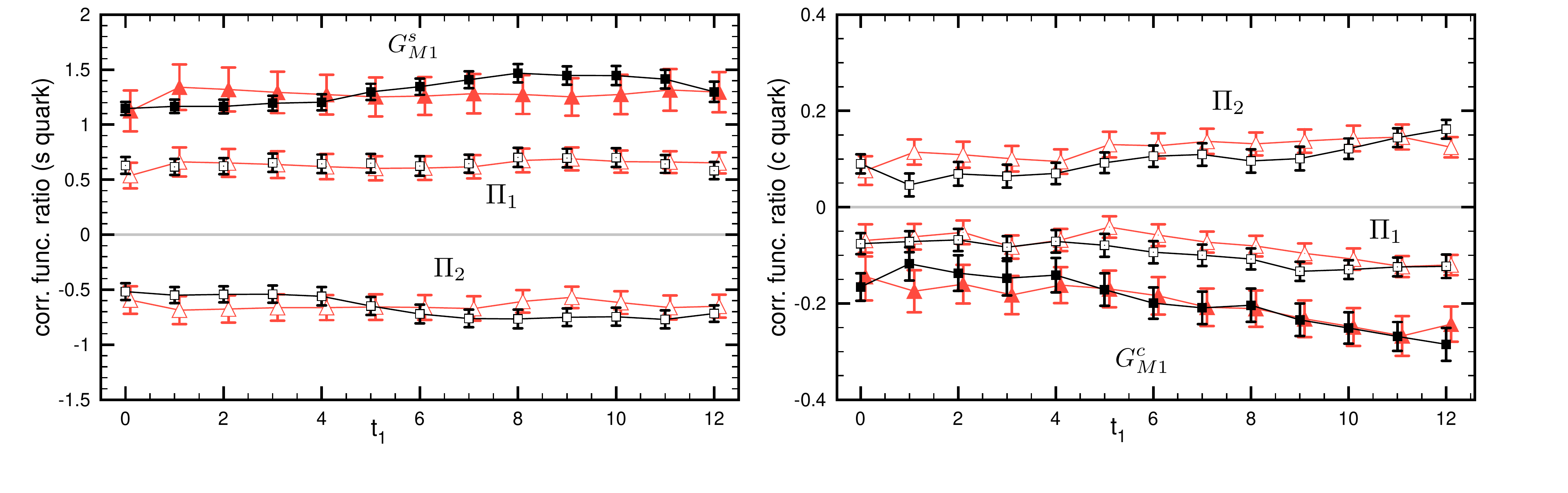}
	\caption{\label{plato} The correlation function ratios $\Pi_1$ and $\Pi_2$ in Eq.~\eqref{pi1pi2} as functions of the current insertion time ($t_1$) for $s$- and $c$-quark sectors. We also display $G_M^{s,c}$ obtained using Eq.~\eqref{mffavg}. The squares (triangles) denote the kinematical case when $\omcs$ ($\omc$) at rest.}
\end{figure}

We define the sum of all correlation-function ratios as
\begin{equation}\label{pi1pi2}
	\Pi_1=\frac{C(q^2)}{|\bf{q}|}\frac{1}{6}\sum_{k,l}\Pi_l(q_k,{\bf 0}; \Gamma_k;l), \quad
	\Pi_2=\frac{C(q^2)}{|\bf{q}|}\frac{1}{6}\sum_{k,l}\Pi_k(q_k,{\bf 0}; \Gamma_l;l),
\end{equation}
so that Eq.~\eqref{mff} and Eq.~\eqref{qff} becomes,
\begin{align}
	\begin{split}\label{mffavg}
	G_{M1}(q^2)&=\Pi_1-\frac{m_\ast}{E_\ast}\Pi_2, 
	\end{split}\\
	\begin{split}\label{qffavg}
	G_{E2}(q^2)&=\Pi_1+\frac{m_\ast}{E_\ast}\Pi_2.
	\end{split}
\end{align}
Fig.~\ref{plato} illustrates the $\Pi_1$ and $\Pi_2$ as functions of the current insertion time, $t_1$, for $s$- and $c$-quark sectors separately. The two ratios have opposite sign and they add constructively when they are subtracted. We extract the form factors by fitting the correlation-function ratios by a horizontal line where a plateau develops. We illustrate both kinematical cases giving consistent results within their error bars. A clear plateau can be realized in both kinematical cases, being more flat when $\omc$ is produced at rest. We fit the correlation function ratios in the range $t_1=[3,6]$. The statistical errors, on the other hand, are smaller when $\omcs$ is at rest. The values of the form factors from the two kinematical cases are consistent with each other.

It is straightforward to extract $G_{E2}$ once we construct the correlation function ratios for $G_{M1}$. The correlation functions have opposite signs and are of similar magnitudes, which result in a vanishing value for $G_{E2}$ when they are added. We determine $G_{E2}$ by fitting $\Pi_1$ and $\Pi_2$ separately and combining the results. This procedure gives consistent results with fitting the sum of the correlation ratios. 

%%%%%%%%%%%%%%%%%%%%%% Figure 1  %%%%%%%%%%%%%%%%%%%%%%
% %%%%%%%%%%%%%%%%%%%%%%%%%%%%%%%%%%%%%%%%%%%%%%%%%%%%%%%%%%%%%%%%%%

Our numerical results are reported in Table~\ref{formf}. We give the values of $G_{M1}$ and $G_{E2}$ form factors at the lowest allowed momentum transfer and at zero momentum transfer for the two kinematical cases as explained above. The quark sector contributions to each form factor are given separately. The form factors can be inferred from individual quark contributions by
\begin{equation}~\label{comb}
	G_{M1}(Q^2)=\frac{2}{3}\times G^c_{M1}(Q^2)-\frac{1}{3} \times G^s_{M1}(Q^2),
\end{equation}
and similarly for $G_{E2}(Q^2)$. The values of the form factors at $Q^2=0$ are extracted using the scaling assumption in Eq.~\eqref{scaling}. 

Similarly to what has been observed in the case of elastic form factors~\cite{Can:2013tna}, $M1$ form factor is dominantly determined by the contribution of the $s$-quark sector, which is approximately one order of magnitude larger than that of the $c$-quark sector. This pattern is consistent with hyperon transition form factors~\cite{Leinweber:1992pv}: The heavier quark contribution is systematically smaller than that of the light quarks. From a quark-model point of view, the coupling of the photon to the light quarks prevails in the heavy-quark limit and the heavy quark acts as a spectator. In this limit, the transition proceeds dominantly through the spin flip of the light degrees of freedom and only $M1$ transition is allowed. Only finite mass effects of the heavy quark may lead to a nonzero value of $E2$ form factor. Our results show that the two quark sectors contribute with opposite signs and yield a value with a statistical error of approximately $5\%$ when combined via Eq.~\eqref{comb}. The values from the two kinematical cases are consistent with each other within their error bars. 

%%%%%%%%% table of results %%%%%%%%%%%%%%%%%%%%%%
\begin{table}[t]
	\caption{ The results for $G_{M1}$ and $G_{E2}$ form factors at the lowest allowed four-momentum transfer and at zero momentum transfer for the two kinematical cases. The quark sector contributions to each form factor are given separately. Note that the statistical uncertainty is large in $G_{E2}$ and results are consistent with zero. 
}
%	\addtolength{\tabcolsep}{6pt}
\begin{center}
	{
	\setlength{\extrarowheight}{7pt}
\begin{tabular*}{\textwidth}{@{\extracolsep{\fill}}cc|ccc|ccc}
			\hline\hline 
			\multicolumn{2}{c}{$Q^2$[GeV$^2$]} & $G^s_{M1}(Q^2)$ & $G^c_{M1}(Q^2)$ & $G_{M1}(Q^2)$ & $G^s_{E2}(Q^2)$ & $G^c_{E2}(Q^2)$ & $G_{E2}(Q^2)$  \\
			\hline
			 \multirow{2}{*}{$\omcs$ at rest} & 0.180 & 1.257(67) & -0.167(33) & -0.530(28)  & 0.041(132) & 0.008(26) & -0.008(50)\\
			 								 & 0     & 1.622(87) & -0.175(34) & -0.657(33)  & 0.052(171) & 0.009(27) & -0.012(62) \\
			 \multirow{2}{*}{$\omc$ at rest}  & 0.168 & 1.269(177) & -0.174(37) & -0.539(78) & -0.035(124) & 0.061(25) & 0.052(48) \\
			 								 & 0     & 1.637(229) & -0.183(39) & -0.667(96) & -0.045(160) & 0.064(27) & 0.058(60)  \\
			\hline \hline
		
\end{tabular*}
	\label{formf}
	}
\end{center}
\end{table}
%%%%%%%%%%%%%%%%%%%%%%%%%%%%%%%%%%%%%%%%%%%%%%%%%%%%%%%%%%%%%%%%%%

In contrast, the extracted values of $G_{E2}$ at finite and zero momentum transfer are small and consistent with zero within their error bars. A comparison of $G_{M1}$ and $G_{E2}$ reveals that the transition is entirely determined by $M1$ transition. In quark model, the quadrupole transition moments arise from the tensor-induced D-state admixtures of the single-quark wavefunctions~\cite{Isgur:1981yz} and the two-quark exchange currents~\cite{Buchmann:1991cy, Buchmann:1996bd}. In the first, the spins of the quarks remain the same but an S-state quark is changed into a D-state. The latter can be interpreted as the spin flip of a diquark inside the baryon. Given the dependence of the tensor force on the inverse quark mass, one would expect to obtain a smaller $G_{E2}$ value for heavy baryons as compared to that in the light-baryon sector, in consistency with what we have found. The smallness of the $E2$ form factor can also be understood as a chiral suppression. The $E2$ amplitude is dominated by pion loops and the leading contribution comes from chiral logs which can be computed in heavy-baryon chiral perturbation theory~\cite{Butler:1993ht, Savage:1994wa}.

The Sachs form factors calculated above can be related to phenomenological observables such as the helicity amplitudes and the decay width. The relation between the Sachs form factors extracted in this work and the standard definitions of electromagnetic transition amplitudes $f_{M1}$ and $f_{E2}$ in the rest frame of $\omcs$ are given by~\cite{Nozawa:1989pu, Sato:2000jf}
\begin{align}
		f_{M1}(q^2)&=\frac{\sqrt{4\pi\alpha}}{2m}\left(\frac{|{\bm{q}}|m_\ast}{m}\right)^{1/2}\frac{G_{M1}(q^2)}{[1-q^2/(m+m_\ast)^2]^{1/2}},\\
		f_{E2}(q^2)&=\frac{\sqrt{4\pi\alpha}}{2m}\left(\frac{|{\bm{q}}|m_\ast}{m}\right)^{1/2}\frac{G_{E2}(q^2)}{[1-q^2/(m+m_\ast)^2]^{1/2}},
\end{align}
where $\alpha=1/137$ is the fine structure constant. The helicity amplitudes $A_{1/2}$ and $A_{3/2}$ can be deduced from the transition amplitudes as follows:
	\begin{align}
			A_{1/2}(q^2)&=-1/2[f_{M1}(q^2)+3f_{E2}(q^2)],\\
			A_{3/2}(q^2)&=-\sqrt{3}/2[f_{M1}(q^2)-f_{E2}(q^2)].
	\end{align}
Then the decay width is given by~\cite{Agashe:2014kda}
	\begin{equation}
		\Gamma=\frac{m_\ast m}{8\pi}\left(1-\frac{m^2}{m_\ast^2}\right)^2\{|A_{1/2}(0)|^2+|A_{3/2}(0)|^2\},
	\end{equation} 
where we have used the constraint ${\bf q}=(m_\ast^2-m^2)/2m_\ast$ at $q^2=0$. The decay width can also be obtained from the Sachs form factors:
	\begin{equation}
		\Gamma=\frac{\alpha}{16}\frac{(m_\ast^2-m^2)^3}{m^2 m_\ast^3}\{3 |G_{E2}(0)|^2+|G_{M1}(0)|^2\}.
	\end{equation}

%%%%%%%%% table of results %%%%%%%%%%%%%%%%%%%%%%
\begin{table}[t]
	\caption{ The results for the helicity amplitudes and the decay width in the rest frame of $\omcs$. The helicity amplitudes are given at finite and zero momentum transfer. The zero-momentum values are obtained using the scaling assumption in Eq.~\eqref{scaling}.
}
%	\addtolength{\tabcolsep}{6pt}
\begin{center}
	{
	\setlength{\extrarowheight}{7pt}
\begin{tabular*}{0.8\textwidth}{@{\extracolsep{\fill}}cccccccc}
			\hline\hline 
			  $Q^2$ & $f_{M1}$ & $f_{E2}$ & $A_{1/2}$ & $A_{3/2}$ & $\Gamma$ \\
			  	\scriptsize{[GeV$^2$]}	& \scriptsize{$10^{-2}$[GeV$^{-1/2}$]} & \scriptsize{$10^{-2}$[GeV$^{-1/2}$]} & \scriptsize{$10^{-2}$[GeV$^{-1/2}$]} & \scriptsize{$10^{-2}$[GeV$^{-1/2}$]} & \scriptsize{[keV]} \\
			\hline
				0.180 & -0.795(42) & -0.012(75) & 0.416(116) & 0.678(71) &  \\
			0 & -0.988(50) & -0.018(93) & 0.521(145) & 0.840(88) & 0.074(8) \\
			\hline

\end{tabular*}
	\label{phen}
	}
\end{center}
\end{table}
%%%%%%%%%%%%%%%%%%%%%%%%%%%%%%%%%%%%%%%%%%%%%%%%%%%%%%%%%%%%%%%%%%
	
Since the above formulas are continuum relations, we use the experimental values of $\omc$ and $\omcs$ masses in calculating the helicity amplitudes and the decay width. Our numerical results for the helicity amplitudes in the rest frame of $\omcs$ and the decay width, at finite and zero momentum transfer, are reported in Table~\ref{phen}. A comparison to the $N\gamma \rightarrow \Delta$ transition~\cite{Agashe:2014kda} reveals that, the helicity amplitudes are suppressed roughly by five orders of magnitude due to diminishing contribution of the heavy quark, the overall reduction in the transition form factors and the larger baryon masses. 

Since no strong decay channel is kinematically allowed, the total decay rate of $\omcs$ is almost entirely in terms of the photon decay mode. Eventually a significantly suppressed value of the $\omcs$-baryon decay width is yielded, making $\omcs$ one of the longest living spin-3/2 charmed hadrons. The suppression in the decay width can be mainly attributed to the small $\omcs$-$\omc$ mass splitting. The decay width in Table~\ref{phen} is translated into a lifetime of $\tau=1/\Gamma=8.901(913)\times 10^{-18}$~sec. 

The electromagnetic transitions of charmed baryons have also been studied within heavy hadron chiral perturbation theory~\cite{Cheng:1992xi, Banuls:1999br} and quark models~\cite{Dey:1994qi, Ivanov:1996fj, Ivanov:1999bk}. It has been found that the charmed baryon electromagnetic decays are suppressed, in qualitative agreement with our result. Of special interest is the $\Sigma^{\ast,+}_c\rightarrow \Sigma^+_c \gamma$ decay having a similarly small width in the quark model~\cite{Ivanov:1999bk}. An enhanced width is foreseen in the $\Sigma^{\ast +}_c\rightarrow \Lambda^+_c \gamma$ decay, which would be interesting to study on the lattice. The literature on $\oog$ transition is limited. Non-relativistic quark model prediction for $\omcs$ decay width~\cite{Dey:1994qi} is one order of magnitude larger than what we have calculated in this work. Note that given the small $\omcs$-$\omc$ mass splitting, such a large width would require a $G_{M1}$ value as large as that of $N\gamma \rightarrow \Delta$ transition. This cannot be justified as we have found that the heavy-quark contribution diminishes and there is no indication that the light quark contribution is enhanced.

In conclusion, we have computed the $\oog$ transition in lattice QCD. The dominant contribution is due to the magnetic dipole form factor, which we have calculated with a statistical precision of about $5\%$. The electric quadrupole transition has been found to be negligibly small in consistency with the quark model. We have extracted the helicity amplitudes and the decay width, which have been found to be suppressed. This transition is of particular interest because of its relevance to current and proposed experimental facilities such as LHCb, PANDA, Belle II, BESIII and J-PARC, which are expected to measure the electromagnetic decay widths of charmed baryons with a higher precision. 

\acknowledgments
Part of numerical calculations in this work were performed on National Center for High Performance Computing of Turkey (Istanbul Technical University) under project number 10462009. The unquenched gauge configurations employed in our analysis were generated by PACS-CS collaboration~\cite{Aoki:2008sm}. We used a modified version of Chroma software system~\cite{Edwards:2004sx}. This work is supported in part by The Scientiﬁc and Technological Research Council of Turkey (TUBITAK) under project number 114F261 and in part by KAKENHI under Contract Nos. 25247036 and 24250294. This work is also supported by the Research Abroad and Invitational Program for the Promotion of International Joint Research, Category (C) and the International Physics Leadership Program at Tokyo Tech.

\end{document}